\documentclass[12pt,a4paper,final]{iopart}

\usepackage{iopams}  
\usepackage{graphicx}
\usepackage[breaklinks=true,colorlinks=true,linkcolor=blue,urlcolor=blue,citecolor=blue]{hyperref}

\begin{document}

\title[Quantum speed limit time  in the presence of disturbance]{Quantum speed limit time  in the presence of disturbance}

\author{S. Haseli$^{1}$}
\address{$^1$Faculty of Physics, Urmia University of Technology, Urmia, Iran.}
\ead{soroush.haseli@gmail.com}
\author{S. Salimi$^{2}$, H. Dolatkhah$^{2}$, A. S. Khorashad$^{2}$}
\address{$^2$Department of Physics, University of Kurdistan, P.O.Box 66177-15175, Sanandaj, Iran}

\begin{abstract}
Quantum theory sets a bound on the minimal time evolution between initial and target states. This bound  is called as quantum speed limit time. It is used to quantify maximal speed of quantum evolution. The  quantum evolution will be faster,  if quantum speed limit time decreases. In this work, we study the quantum speed limit time of a quantum state in the presence of disturbance effects in an environment. We use the model which is provided by Masashi Ban in \href{https://doi.org/10.1103/PhysRevA.99.012116}{Phys. Rev. A 99, 012116 (2019)}. In this model two quantum systems $\mathcal{A}$ and $\mathcal{S}$  interact with environment sequentially. At first, quantum system $\mathcal{A}$ interacts with the environment $\mathcal{E}$ as an auxiliary system then quantum system $\mathcal{S}$ interacts with disturbed environment immediately. In this work, we consider dephasing coupling with two types of environment with different spectral density: Ohmic and Lorentzian.  We observe that, non-Markovian effects will be appear in the dynamics of quantum system $\mathcal{S}$ by the interaction of quantum system $\mathcal{A}$ with the environment. Given the fact that quantum speed limit time reduces due to non-Markovian effects, we show that disturbance effects will reduce the quantum speed limit time.

\end{abstract}

\vspace{2pc}

\section{Introduction}	
Quantum speed limit QSL time determines the speed of the quantum evolution for the dynamics of quantum systems. It sets a bound on the minimal evolution time needed for a quantum state of a closed or open quantum system to evolve from initial state to target state. It has many important applications in the field of quantum physics ranging from, quantum metrology \cite{Giovanetti}, computation \cite{Lloyd}, communication \cite{Bekenstein} to nonequilibrium quantum thermodynamics \cite{Deffner}   and quantum optimal control \cite{Caneva}. In Ref. \cite{Mandelstam},  Mandelstam and Tamm MT have provided a new insight on energy-time uncertainty relation. They showed that, the QSL time $\tau_{QSL}$ for the  closed quantum systems is given by  
\begin{equation}\label{MT}
\tau \geq \tau_{QSL}=\frac{\pi \hbar}{2 \Delta E},
\end{equation}
where $\Delta E = \sqrt{\langle \hat{H}^{2} \rangle - \langle \hat{H} \rangle^{2}}$ is the inverse of the variation of energy of the initial state and $\hat{H}$ is the time-independent Hamiltonian  describing the dynamics of quantum system. Also in Ref. \cite{Margolus}, Margolus and Levitin ML introduced the QSL time for the closed quantum systems based on the mean energy $E=\langle \hat{H} \rangle$ due to the ground state as 
\begin{equation}\label{ML}
\tau \geq \tau_{QSL}=\frac{\pi \hbar}{2  E}.
\end{equation}
Considering and combining the results of  MT and ML bounds, the QSL time for closed quantum systems and orthogonal states can be expressed as 
\begin{equation}
\tau \geq \tau_{QSL}=\max \lbrace  \frac{\pi \hbar}{2 \Delta E} , \frac{\pi \hbar}{2  E} \rbrace.
\end{equation}
Due to the fact that the Hamiltonian is the generator of unitary evolution, it is reasonable to express  the QSL time based on the initial energy of the system. In Refs.\cite{Giovannetti1,Jones,Zwierz,Deffner,Pfeifer,Pfeifer1}, the QSL time for closed quantum system and orthogonal states is generalized to non-orthogonal states and driven systems. 

In the real world the interaction of the system with its surrounding is inevitable. So, the theory of open quantum systems is used to examine such systems \cite{Breuer}. Due to the direction of information flow, the dynamics of open quantum systems can be classified into Markovian and non-Markovian quantum evolution. In Markovian dynamics  the information only flow from the system to
the environment, i.e the system smoothly loses its information. For non-Markovian dynamics the information flow-back from the environment to the system  in some moments during quantum evolution.

In recent years, the QSL time for open quantum systems has been widely studied. For open quantum systems, QSL time has characterized using quantum Fisher information  \cite{Taddei,Escher}, Bures angle \cite{Deffner1}, relative purity \cite{del,Zhang} and other proper distance measures \cite{Xu,Mondal,Levitin,Xu1,Meng}. In Ref. \cite{del} del Campo et al.  showed that when  the master equation has the Lindblad form the relative purity bound of QSL time is  similar to the MT bound. For pure initial state, Deffner and Lutz employed Bures angle to introduce QSL time and they have defined a unique bound that covers MT and ML bound \cite{Deffner1}. 

In addition to all these attempts, various studies have also been done in the context of the QSL time for open quantum systems such as the connection between initial state and QSL time  \cite{Wu,xiong,Zhang1}, relativistic effects on QSL time\cite{Salman,Haseli}, QSL time  based on alternative fidelity  \cite{Ektesabi}, QSL time in non-equilibrium environment \cite{Cai}, QSL time  for multipartite open quantum systems \cite{Liu}. 
 
In this work, we consider an interesting case of open quantum systems in which the quantum system interacts with disturbed environment. We review the model of two quantum systems that interact with environment sequentially. First, one of the quantum systems interacts with initial non-disturbed environment in a finite time and disturbs the environment, afterwards the second system interacts with the disturbed environment  \cite{Ban}. In this work we consider the case in which the two quantum system interacts with common Bosonic environment through a dephasing coupling consecutively. The quantum evolution of the second quantum system can be non-Markovian. It is due to the disturbance of the environment caused by interacting of the first quantum system with initial environment. Therefore, it can be concluded that,  even if the evolution of the first quantum system is Markovian, the dynamics of the second quantum system can be non-Markovian.  We will quantify the degree of non-Markovianity stems from disturbance in terms of environmental parameter. Deffner and Lutz have showed that back-flow of information from environment to quantum system, i.e. non-Markovian effects will increase the speed of quantum evolution and hence will reduce the QSL time \cite{Deffner1}. So we expect that the disturbance of the environment, due to  its interaction with the first quantum system, will reduce the QSL time for quantum evolution of the second quantum system. Here, We will consider the Ohmic and Lorentzian spectral density of the environment. It is worth noting that the amount of disturbance in the environment depends on the state of the first quantum system. We will show that the disturbance is strong when the state of the first quantum system   which disturbs the environment has zero coherence and the disturbance is weak when it has maximum value of coherence \cite{Masashi1}. 

In this work, we review and use the relative purity bound of QSL time for arbitrary initial states \cite{Zhang}. The motivation to use this bound is that it can be used for any arbitrary initial state, whether pure or mixed. In Ref. \cite{Zhang}, Zhang et al.  showed that the QSL time for dephasing model is depend on  the quantum coherence of the initial state.

This work is organized as follows. In Sec. \ref{Sec2}, first we review the dynamics of open quantum system interacts with environment with disturbance.  Then we provide a general formula for the dynamic of a two-level quantum system interacting with disturbed environment. We consider  environments with Ohmic and Lorentzian spectral density. We also quantify the degree of non-Markovianity which is arises due to diturbance by using $l_1$-norm of quantum coherence. In Sec. \ref{Sec2}, we calculate QSL time for the dynamics of open quantum system in the the presence of disturbance and compare our results with the case there exist no disturbance effects. The conclusion and summary of this work is given in Sec.\ref{Sec4}.
\section{Dynamics of open quantum system interacting with disturbed environment}\label{Sec2}
Here, we consider two quantum systems $\mathcal{A}$ and $\mathcal{S}$. At first,  system $\mathcal{A}$ interacts with the environment $\mathcal{E}$ as an auxiliary system from time $t_0$ to time $t_1$ and leads to disturbances in the environment . Then system $\mathcal{S}$ interacts with disturbed environment from time $t_2$ to time $t_3$. For that time ordering from $t_0$ to $t_3$, the inequality holds in the form $t_0 <  t_1 \leq t_2 < t_3$. If the time interval $t_2-t_1$ between the end of the interaction of system $\mathcal{A}$ with the environment and the start of the interaction of system $\mathcal{S}$ with the disturbed environment is greater than the correlation time of the environment then the disturbance effects will be ignored.

Let us supposed that the two quantum system $\mathcal{A}$ and $\mathcal{S}$ are two-level systems. We also assume that  $\mathcal{A}$ and $\mathcal{S}$ interact with Bosonic environment through a dephasing model. The Hamiltonian of the two qubit quantum system  and its interaction Hamiltonian are given by $H_{i}=\hbar \omega_{i} \sigma_{z}^{i}/2$ and $H_{i \mathcal{E}}=\hbar/2 \sigma_{z}^{i} \otimes \mathcal{E}$ respectively, where $\mathcal{E}=\sum_{k}g_{k} (a_{k}+a_{k}^{\dag})$ is the  environmental operator, $i=\mathcal{A},\mathcal{S}$ and $\sigma_{z}^{i}$ is the $z$-component of the Pauli operator, $a_k$($a_{k}^{\dag}$) is  annihilation(creation) operator of the $k$-th environmental oscillator with angular frequency $\omega_k$. Hamiltonian of the environment $\mathcal{E}$ given in the form $H_\mathcal{E}=\sum_{k}\hbar \omega_{k} a_{k}^{\dag}a_{k}$. The whole system consists of $\mathcal{A}$, $\mathcal{S}$, and $\mathcal{E}$, from $t_0$ to $t_1$ are evolved through Hamiltonian $H_{\mathcal{A}}+H_{\mathcal{E}}+H_{\mathcal{A}\mathcal{E}}$ and from $t_2$ to $t_3$ are evolved through Hamiltonian $H_{\mathcal{S}}+H_{\mathcal{E}}+H_{\mathcal{S}\mathcal{E}}$. The disturbance effects are important for us, So we assume that $t_1=t_2$. The evaluated state of the whole system is given by
\begin{equation}
\rho_{\mathcal{A}\mathcal{S}\mathcal{E}}^{out}=e^{({L_\mathcal{S}+L_\mathcal{E}+L_\mathcal{SE}})(t_3-t_2)}e^{({L_\mathcal{A}+L_\mathcal{E}+L_\mathcal{AE}})(t_1-t_0)}\rho_{\mathcal{A}\mathcal{S}\mathcal{E}}^{in},
\end{equation}  
where $L_{o} (\square)=-i/\hbar \left[ H_{o}, \square \right] $ with $ o \in \left\lbrace  \mathcal{A}, \mathcal{S}, \mathcal{E}, \mathcal{AE}, \mathcal{SE} \right\rbrace $ is Liouvillian superoperator \cite{Kubo,Weiss}.

We assume that there exist no initial correlation between quantum systems $\mathcal{A}$ and $\mathcal{S}$, i.e. $\rho_{\mathcal{A}\mathcal{S}}^{in}=\rho_{\mathcal{A}}^{in} \otimes \rho_{\mathcal{S}}^{in}$. Let us consider the initial state of two quantum systems $\mathcal{A}$ and $\mathcal{S}$ as  
\begin{eqnarray}
\rho_{\mathcal{A}}^{in}=\rho_{ee}^{a}\vert e \rangle\langle e \vert +\rho_{eg}^{a}\vert e \rangle\langle g \vert  +\rho_{ge}^{a}\vert g \rangle\langle e \vert +\rho_{gg}^{a}\vert g \rangle\langle g \vert,  \\
\rho_{\mathcal{S}}^{in}=\rho_{ee}^{s}\vert e \rangle\langle e \vert +\rho_{eg}^{s}\vert e \rangle\langle g \vert  +\rho_{ge}^{s}\vert g \rangle\langle e \vert +\rho_{gg}^{s}\vert g \rangle\langle g \vert,  
\end{eqnarray}
shuch that $ \sigma_z \vert e \rangle= \vert e \rangle$ and $\sigma_z \vert g \rangle=- \vert g \rangle$.
Using the method outlined in Ref.\cite{Ban}, one can derive the reduced time evolution of two quantum systems $\mathcal{A}$ and $\mathcal{S}$ as
\begin{eqnarray}
\label{g_4}
\rho_{\mathcal{A}}(t)&=& \rho_{ee}^{a} \vert e \rangle\langle e \vert +  \rho_{gg}^{a} \vert g \rangle\langle g \vert +\rho_{eg}^{a} e^{-i \omega_{a}(t-t_0)-\mathcal{G}_{\mathcal{R}}(t,t_0)} \vert e \rangle\langle g \vert
 \nonumber \\
&+&\rho_{ge}^{a} e^{i \omega_{a}(t-t_0)-\mathcal{G}_{\mathcal{R}}(t,t_0)} \vert g \rangle\langle e \vert, \quad 0 \leq t \leq t_1, \\
\label{g_5}
\rho_{\mathcal{S}}(t)&=& \rho_{ee}^{s} \vert e \rangle\langle e \vert +  \rho_{gg}^{s} \vert g \rangle\langle g \vert 
 \nonumber \\
&+&\rho_{eg}^{s} f(t)  \vert e \rangle\langle g \vert + \rho_{ge}^{s}  f^{*}(t) \vert g \rangle\langle e \vert, \quad t_2=t_1 \leq t \leq t_3, 
\end{eqnarray}
with 
\begin{eqnarray}\label{g_6}
f(t)&=&\left[ \cos \mathcal{G}_{\mathcal{I}}(t,t_2;t_1,t_0)-i \langle \sigma_{z}^{a} \rangle \sin \mathcal{G}_{\mathcal{I}}(t,t_2;t_1,t_0) \right] \nonumber \\
&\times & e^{i \omega_{s}(t-t_2)-\mathcal{G}_{\mathcal{R}}(t,t_2)},
\end{eqnarray}
Here $ \langle \sigma_{z}^{a} \rangle =tr_{\mathcal{A}}(\rho_{\mathcal{A}}(0)\sigma_{z}^{a})$ and 
\begin{eqnarray}
\mathcal{G}_{\mathcal{R},\mathcal{I}}(t,t_k)=\int_{t_{k}}^{t}d \tau \int_{t_{k}}^{\tau}d  \tau^{\prime} \mathcal{C}_{\mathcal{R},\mathcal{I}}(\tau - \tau^{\prime}), \quad k=0,2 \nonumber \\
\mathcal{G}_{\mathcal{R},\mathcal{I}}(t,t_2;t_{1},t_{0})=\int_{t_{2}}^{t}d \tau \int_{t_{0}}^{t_1}d  \tau^{\prime} \mathcal{C}_{\mathcal{R},\mathcal{I}}(\tau - \tau^{\prime}), \qquad
\end{eqnarray}
in above equations, $\mathcal{C}_{\mathcal{R}}(\tau - \tau^{\prime})$ and  $\mathcal{C}_{\mathcal{I}}(\tau - \tau^{\prime})$ are the real and imaginary parts of the two-time correlation function 
\begin{eqnarray}
\mathcal{C}(\tau - \tau^{\prime})&=&\mathcal{C}_{\mathcal{R}}(\tau - \tau^{\prime})+i \mathcal{C}_{\mathcal{I}}(\tau - \tau^{\prime}) \nonumber \\
&=&\langle \mathcal{E}(\tau \vert t_{0})\mathcal{E}(\tau^{\prime} \vert t_{0} ) \rangle_{\mathcal{E}}  \\
&=& \sum_{k}g_{k}^{2}\left\lbrace (\bar{N}_{k}+1)e^{-i\omega_{k}(\tau-\tau^{\prime})} + \bar{N}_{k}e^{i\omega_{k}(\tau-\tau^{\prime})}\right\rbrace, \nonumber 
\end{eqnarray}
where $\langle \bullet \rangle_{\mathcal{E}} = tr_{\mathcal{E}}\left[ \bullet \rho_{\mathcal{E}} \right] $, $\bar{N}_{k}=\langle a_{k}^{\dag}a_{k}\rangle_{\mathcal{E}}$ and
\begin{eqnarray}
\mathcal{E}(\tau \vert t_{0})&=&e^{i H_{\mathcal{E}}(\tau - t_0)/\hbar}\mathcal{E}e^{-i H_{\mathcal{E}}(\tau - t_0)/\hbar}  \\
&=&\sum_{k}g_{k}( e^{-i \omega_{k}(\tau-t_{0})/\hbar}a_{k} + e^{i \omega_{k}(\tau-t_{0})/\hbar}a_{k}^{\dag}),\nonumber \\ 
\mathcal{E}(\tau^{\prime} \vert t_{0})&=&e^{i H_{\mathcal{E}}(\tau^{\prime} - t_0)/\hbar}\mathcal{E}e^{-i H_{\mathcal{E}}(\tau^{\prime} - t_0)/\hbar}  \\
&=&\sum_{k}g_{k}( e^{-i \omega_{k}(\tau^{\prime}-t_{0})/\hbar}a_{k} + e^{i \omega_{k}(\tau^{\prime}-t_{0})/\hbar}a_{k}^{\dag}).\nonumber
\end{eqnarray}
In general, for dephasing coupling with Bosonic environment  we have 

\begin{eqnarray}
\label{g_1}
\mathcal{G}(t,t_m)&=&\mathcal{G}_{\mathcal{R}}(t,t_m)+i\mathcal{G}_{\mathcal{R}}(t,t_m)  \\
&=&\sum_{k}(\frac{g_{k}}{\omega_{k}})^{2}(1-e^{-i\omega_{k}(t-t_m)}+i\omega_{k}(t-t_m)) \nonumber \\
&=& \int_{0}^{\infty}d\omega J(\omega)\frac{1-e^{-i\omega(t-t_m)}+i\omega(t-t_m)}{\omega^{2}}, \quad m=0,2 \nonumber \\ 
 \label{g_2}
\mathcal{G}(t,t_2;t_{1},t_{0})&=&\mathcal{G}_{\mathcal{R}}(t,t_2;t_{1},t_{0})+i\mathcal{G}_{\mathcal{I}}(t,t_2;t_{1},t_{0})\\ 
&=&\sum_{k}(\frac{g_{k}}{\omega_{k}})^{2}(e^{-i\omega_{k}(t-t_1)}-e^{-i\omega_{k}(t_2-t_1)}-e^{-i\omega_{k}(t-t_0)}+e^{-i\omega_{k}(t_2-t_0)})\nonumber \\
&=&\int_{0}^{\infty}d\omega J(\omega)\frac{e^{-i\omega(t-t_1)}-e^{-i\omega(t_2-t_1)}-e^{-i\omega(t-t_0)}+e^{-i\omega(t_2-t_0)}}{\omega^{2}}, \nonumber
\end{eqnarray} 
where $J(\omega)$ is the spectral density of the environment. In the following we will consider dephasing model with Ohmic and Lorentzian spectral density for environment.  
\subsection{Dephasing model with Ohmic spectral density}
In this work, it is assumed that two quantum system interacts with a common Bosonic environment through dephasing coupling. Let us  consider the case in which the environment is initially in the ground state i.e. $\bar{N}_{k}=0$ and it has Ohmic spectral density 
\begin{equation}\label{g_3}
J(\omega)=\eta \frac{\omega^{s}}{\omega_{c}^{s-1}}\exp(-\frac{\omega}{\omega_c}),
\end{equation}
where $\omega_c$ is the cutoff frequency, $s$ is an Ohmicity parameter and $\eta$ is a dimensionless coupling constant.  From Eqs. \ref{g_1}, \ref{g_2} and \ref{g_3},  we obtain

\begin{eqnarray}
\mathcal{G}_{\mathcal{R}}(t,t_m)&=&  \\
&=&\eta \Gamma\left[ s-1 \right]\left( 1- \frac{\cos[(s-1)\arctan[\omega_c (t-t_m)]]}{[1+(\omega_c (t-t_m))^{\frac{s-1}{2}}]}\right) ,  m=0,2, \nonumber \\
\mathcal{G}_{I}(t,t_2;t_1,t_0)&=&\psi(t-t_0)-\psi(t-t_1)-\psi(t_2-t_0)+\psi(t_2-t_1),
\end{eqnarray}
for sub Ohmic ($s<1$) and super Ohmic($s>1$) environment with
\begin{equation}
\psi(t)=\eta \Gamma\left[ s-1 \right]\left(  \frac{\sin[(s-1)\arctan[\omega_c t]]}{[1+(\omega_c t)^{\frac{s-1}{2}}]}\right) .
\end{equation}
and 
\begin{eqnarray}
\mathcal{G}_{\mathcal{R}}(t,t_m)&=&\frac{1}{2}\eta \ln [1+(\omega_c (t-t_m))^{2}], \quad m=0,2, \\
\mathcal{G}_{I}(t,t_2;t_1,t_0)&=&\psi(t-t_0)-\psi(t-t_1)-\psi(t_2-t_0)+\psi(t_2-t_1),
\end{eqnarray}
for Ohmic ($s=1$) environment with
\begin{equation}
\psi(t)=\eta \arctan [\omega_c t].
\end{equation}
From hereafter, we set $t$ for the time elapsed from time
$t_2$ at which quantum system $\mathcal{S}$ starts to interaction with environment, i.e. we set $(t-t_2)\rightarrow t$. Thus from Eqs.\ref{g_5} and \ref{g_6}, the dynamics of reduced quantum system $\mathcal{S}$ can be written as 
\begin{eqnarray}
\label{g_7}
\label{g_8}
\rho_{\mathcal{S}}(t)&=& \rho_{ee}^{s} \vert e \rangle\langle e \vert +  \rho_{gg}^{s} \vert g \rangle\langle g \vert +\rho_{eg}^{s} f(t)  \vert e \rangle\langle g \vert + \rho_{ge}^{s}  f^{*}(t) \vert g \rangle\langle e \vert ,
\end{eqnarray}
with 
\begin{eqnarray}\label{g_8}
f(t)&=&\left[ \cos \mathcal{G}_{\mathcal{I}}(t,t_2;t_1,t_0)-i \langle \sigma_{z}^{a} \rangle \sin \mathcal{G}_{\mathcal{I}}(t,t_2;t_1,t_0) \right] \nonumber \\
&\times & e^{i \omega_{s}(t)-\mathcal{G}_{\mathcal{R}}(t)}.
\end{eqnarray}
For the case in which  there is not exist disturbance effect, we have
\begin{eqnarray}\label{g_8}
f(t)&=& e^{i \omega_{s}(t)-\mathcal{G}_{\mathcal{R}}(t)}.
\end{eqnarray}
\subsection{Dephasing model with Lorentzian spectral density}
As an anothe dephasing model, let us  consider the case in which the environment is initially in the ground state i.e. $\bar{N}_{k}=0$ and it has Lorentzian spectral density spectral density
\begin{equation}
J(\omega)=(\frac{\gamma}{2 \pi})\frac{\lambda^{2}}{(\omega-\delta)^{2}+\lambda^{2}},
\end{equation}
where $\lambda$ defines the spectral width of the coupling, $\gamma$ is coupling constant and $\delta$ is the frequency of the oscillator supported by the environment.  From Eqs. \ref{g_1}, \ref{g_2} and \ref{g_3},  one can obtain
\begin{eqnarray}
\mathcal{G}_{\mathcal{R}}(t,t_m)&=& \\
 &=&\frac{\gamma}{2 \lambda}\frac{1}{1+(\frac{\delta}{\lambda})^{2}}\left(  \lambda (t-t_m)-\frac{1-(\frac{\delta}{\lambda})^{2}}{1+(\frac{\delta}{\lambda})^{2}}(1-e^{\lambda(t-t_m)}\cos \delta t)-\frac{2\frac{\delta}{\lambda}}{1+(\frac{\delta}{\lambda})^{2}}\sin \delta t\right)  ,\nonumber
\end{eqnarray}
\begin{eqnarray}
\mathcal{G}_{I}(t,t_2;t_1,t_0)=\psi(t-t_0)-\psi(t-t_1)-\psi(t_2-t_0)+\psi(t_2-t_1),
\end{eqnarray}

where 

\begin{eqnarray}
&& \\ 
\psi(t)&=&\frac{\gamma}{2 \lambda}\frac{e^{\lambda t}}{1+(\frac{\delta}{\lambda})^{2}} \left( \frac{1-(\frac{\delta}{\lambda})^{2}}{1+(\frac{\delta}{\lambda})^{2}}\sin \delta t + \frac{2\frac{\delta}{\lambda}}{1+(\frac{\delta}{\lambda})^{2}} \cos \delta t \right) . \nonumber
\end{eqnarray}
 We chose $t$ for the time elapsed from time
$t_2$ at which quantum system $\mathcal{S}$ begin to interaction with environment, i.e. we set $(t-t_2)\rightarrow t$. Thus from Eqs.\ref{g_5} and \ref{g_6}, the dynamics of reduced quantum system $\mathcal{S}$ can be written as 
\begin{eqnarray}
\label{g_7}
\rho_{\mathcal{S}}(t)&=& \rho_{ee}^{s} \vert e \rangle\langle e \vert +  \rho_{gg}^{s} \vert g \rangle\langle g \vert +\rho_{eg}^{s} f(t)  \vert e \rangle\langle g \vert + \rho_{ge}^{s}  f^{*}(t) \vert g \rangle\langle e \vert,
\end{eqnarray}
with 
\begin{eqnarray}\label{g_8}
f(t)&=&\left[ \cos \mathcal{G}_{\mathcal{I}}(t,t_2;t_1,t_0)-i \langle \sigma_{z}^{a} \rangle \sin \mathcal{G}_{\mathcal{I}}(t,t_2;t_1,t_0) \right] \nonumber \\
&\times & e^{i \omega_{s}(t)-\mathcal{G}_{\mathcal{R}}(t)}.
\end{eqnarray}
For the case in which  there is not exist disturbance effect, we have
\begin{eqnarray}\label{g_8}
f(t)&=& e^{i \omega_{s}(t)-\mathcal{G}_{\mathcal{R}}(t)}.
\end{eqnarray}
In the following, we study the non-Markovianity of the quantum evolution of quantum system $\mathcal{S}$ 
, which is caused by  the disturbance of environment.
\subsection{Non-Markovianity due to disturbance}
We first review some basic notions of the theory of open quantum systems. The dynamical map is  divisible if it can be written as two completely positive and trace preserving (CPTP) maps $\phi_{t}=\phi_{t,t_p}\phi_{t,0} \quad \forall \quad t_{p} \leq t$. So, the dynamical map is non-divisible  if there exist times $t_p$ at which $\phi_{t,t_p}$ is not (CPTP).  In general, the most important common character of all non-Markovianity measures is that they are introduced based on the non-monotonic time evolution of certain quantities when the divisibility property of (CPTP) maps is violated. We should point out that the inverse statement is not true. Actually, there exist non-divisible dynamical maps that certain quantity shows monotonic behaviour under them.  
\begin{figure}[!] 
\centerline{\includegraphics[scale=0.7]{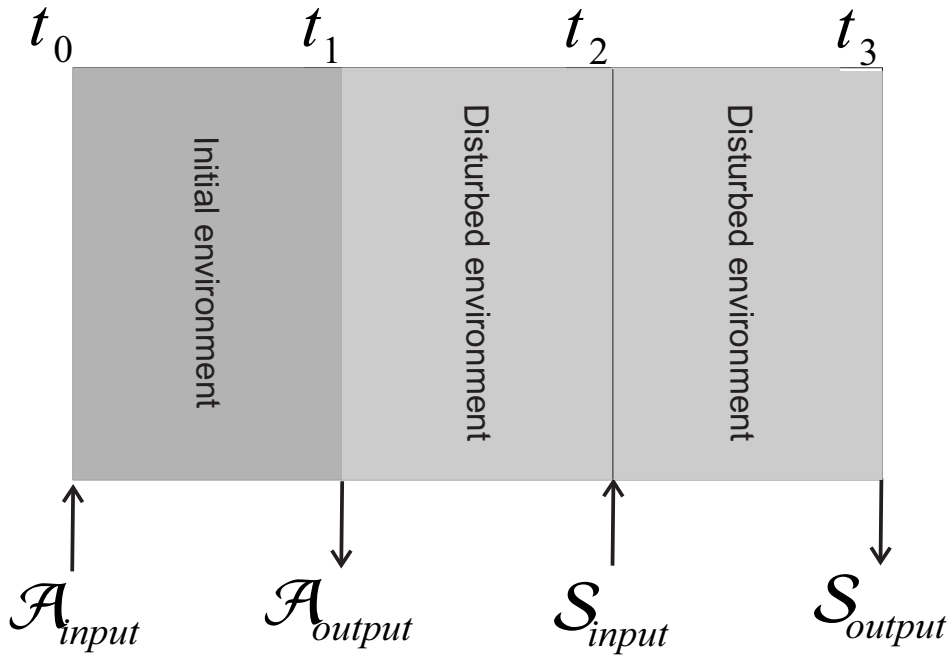}}
\vspace*{8pt}
\caption{The scheme for dynamics of two quantum systems $\mathcal{A}$ and $\mathcal{S}$. Quantum system $\mathcal{A}$ interacts with the environment from time $t_0$ to time $t_1$, then system $\mathcal{S}$ interacts with disturbed environment from time $t_2$ to time $t_3$. }\label{shekl}
\end{figure}
In this work, we focus on non-Markovianity measure which is founded base on the measure of quantum coherence. Quantum coherence is a power full resource in quantum information theory. In recent years significant measure are introduced  to quantify  the quantum coherence, such as relative entropy of coherence \cite{Baumgratz}, trace norm of coherence \cite{Shao} and $l_1$ norme of coherence  \cite{Baumgratz}.  $l_1$-norm of quantum coherence for a quantum state with the density matrixis $\rho$ is defined as \cite{Baumgratz}
\begin{equation}\label{l1norm}
C_{l_1}(\rho)=\sum_{i,j;i\neq j}\vert \rho_{ij} \vert,
\end{equation}
where $\rho_{ij}$'s are the off-diagonal elements of density matrix. When second quantum system $\mathcal{S}$ interacts with disturbed environment, $l_{1}$-norm of coherence is changed as 

\begin{eqnarray}\label{coherence}
C_{l_{1}}(\rho_{\mathcal{S}}(t))&=& \\
&=&2 \vert \rho_{eg} \vert \left( \cos^{2}\mathcal{G}_{I}(t,t_2;t_1,t_0)
+ \langle \sigma_{z}^{a} \rangle^{2}\sin^{2}\mathcal{G}_{I}(t,t_2;t_1,t_0) \right)^{\frac{1}{2}}e^{-\mathcal{G}_{\mathcal{R}}(t)}.\nonumber
\end{eqnarray}

From Eq. \ref{coherence}, one conclouded that when initial state of the quantum system $\mathcal{A}$ is maximally coherent state i,e, $\langle \sigma_{z}^{a} \rangle =0$, disturbance effect has its maximum value while for $\langle \sigma_{z}^{a} \rangle = \pm 1$ disturbance effect has its minimum value.
When quantym dynamical map is incoherent completely positive trace preserding (ICTPT), $l_1$-norm of coherence decreases monotonically. For non-monotonic behavior of $l_1$-norm of coherence, one conclude that the dynamical map is non-divisible and quantum evolution is non-Markovian.  In Ref. \cite{Chanda}, the authors proposed a measure based on $l_1$-norm of coherence  to quantify the degree of non-Markovianity as
\begin{equation}
\mathcal{N}=\max_{\rho_{\mathcal{S}}(0)\in \left\lbrace \vert \psi_{2} \rangle \right\rbrace }\int_{\frac{d C_{l_{1}}(\rho_{\mathcal{S}}(t))}{dt}>0} \frac{d C_{l_{1}}(\rho_{\mathcal{S}}(t))}{dt} dt,
\end{equation}
where the optimization is done over the set of all maximally coherent states $ \vert \psi_{2} \rangle =\frac{1}{\sqrt{2}}\sum_{i=1}^{2}e^{i \varphi_{i}} \vert i \rangle $ and $\varphi_{i} \in [0,2 \pi)$.
\subsubsection{Non-Markovianity when environment have Ohmic spectral density}
\begin{figure}[!] 
\centerline{\includegraphics[scale=0.7]{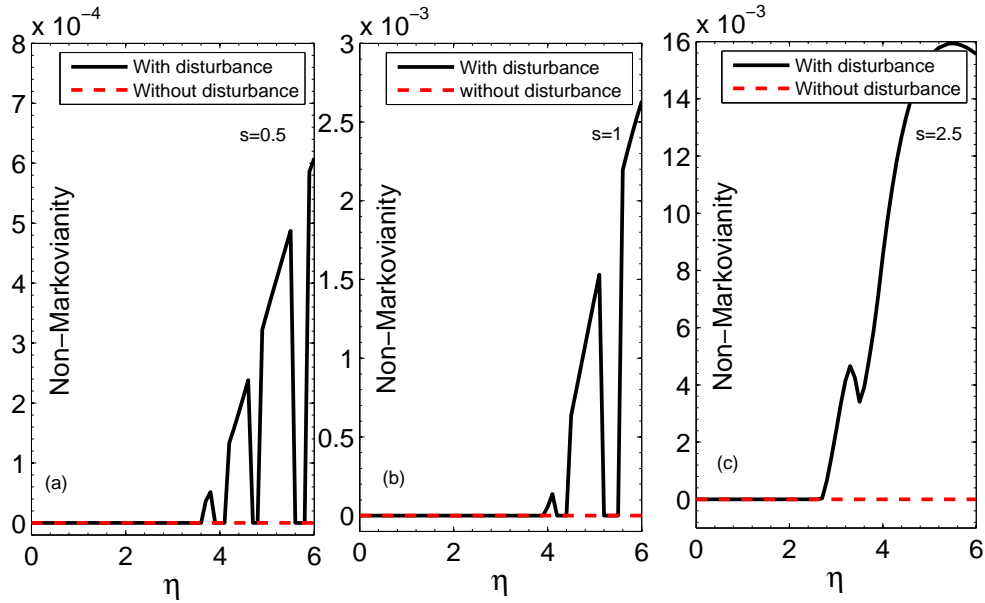}}
\vspace*{8pt}
\caption{(color online)Non-Markovianity as a function of coupling parameter $\eta$ for (a) sub-Ohmic environment with  $s=0.5$. (b)  Ohmic environment with  $s=1$. (d) super-Ohmic environment with  $s=2$. Black line shows the degree of non-Markovianity in the presence of disturbance $\langle \sigma_{z}^{a} \rangle=0.05$ and Red dashed line shows the degree of non-Markovianity without disturbance  . }\label{Fig2}
\end{figure}

Now, we want to investigate how disturbance of environment with ohmic spectral density affects on non-Markovianity. For dephasing model with Ohmic spectral density without disturbance the dynamics is non-Markovian for $s>2$ while it is Markovian for $s \leq 2$. We concentrate on this range in which the dynamic is Markovian without disturbance.  Non-Markovianity for different value of Ohmicity parameter is plotted in Fig. \ref{Fig2}. In Fig.\ref{Fig2}(a), non-Markovianity has been represented as function of coupling parameter for sub-Ohmic environment with $s=0.5$. As can be seen disturbance effects leads to non-Markovianity for $\eta \geq 3.6$, however when the disturbance effect is ignored and environment is in equilibrium the degree of non-Markovianity vanishes.
In Fig.\ref{Fig2}(b), non-Markovianity has been plotted as function of coupling parameter for Ohmic environment with $s=1$. As can be seen disturbance leads to non-Markovianity for $\eta \geq 4$, however when the disturbance effect is ignored and environment is in equilibrium the degree of non-Markovianity vanishes.
In Fig.\ref{Fig2}(c), non-Markovianity has been plottedas function of coupling parameter for super-Ohmic environment with $s=2$. As can be seen disturbance leads to non-Markovianity for $\eta \geq 2.8$, however when the disturbance effect is ignored and environment is in equilibrium the degree of non-Markovianity vanishes.
\subsubsection{Non-Markovianity when environment have Lorentzian spectral density}
\begin{figure}[!] 
\centerline{\includegraphics[scale=0.7]{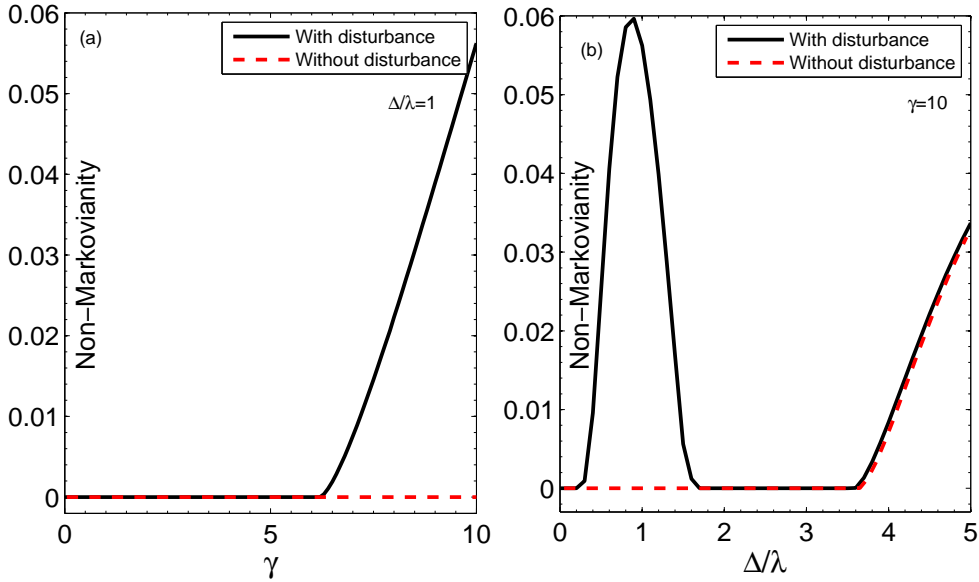}}
\vspace*{8pt}
\caption{(color online) (a) Non-Markovianity as a function of coupling parameter $\gamma$ for Lorentzian environment with  $\Delta / \lambda=1$. (b)Non-Markovianity as a function of  $\Delta / \lambda$ for Lorentzian environment  $\gamma=10$.  Black line shows the degree of non-Markovianity in the presence of disturbance $\langle \sigma_{z}^{a} \rangle=0.05$ and Red dashed line shows the degree of non-Markovianity without disturbance. }\label{Fig3}
\end{figure}

In this part we investigate non-Markovianity for dephasing model with Lorantzian spectral density. In Fig.\ref{Fig3}(a) non-Markovianity has been plottedas a function of coupling parameter $\gamma$. As can be seen in the presence of disturbence dynamic is non-Markovian for $\gamma \geq 6.2$ while in the absence of disturbence the degree of non-Markovianity is zero for all value of $\gamma$. 

 In Fig.\ref{Fig3}(b), non-Markovianity has been plotted as a function of  $\Delta/\lambda$. As can be seen in the presence of disturbence dynamic is non-Markovian around $\Delta/\lambda \simeq 1$ due to disturbance of the environment and $\Delta/\lambda \geq 3.6$ because of pure environmental effects. In the absence of disturbance effect dynamic is non-Markovian for $\Delta/\lambda \geq 3.6$ due to environmental effects.
\begin{figure}[!] 
\centerline{\includegraphics[scale=0.7]{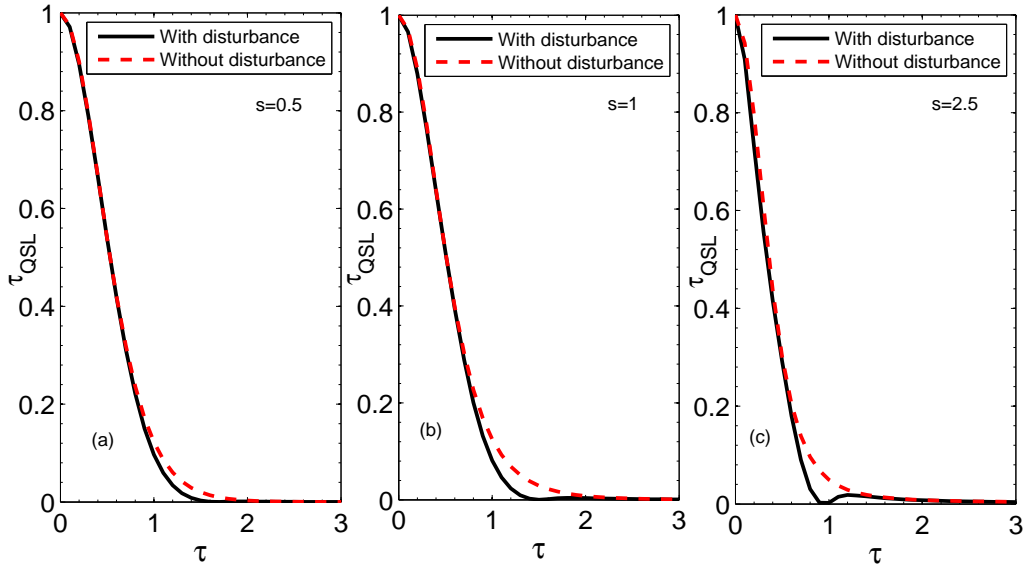}}
\vspace*{8pt}
\caption{(color online) QSL time as a function of initial time parameter $\tau$ for (a) sub-Ohmic environment $s=0.5$, (b)Ohmic environment $s=1$ and (c) Super-Ohmic environment $s=2$, with driving time $\tau_{D}=1$. Black line shows the QSL time in the presence of disturbance $\langle \sigma_{z}^{a} \rangle=0.05$ and Red dashed line shows the QSL time without disturbance. }\label{Fig4}
\end{figure}
%
%
%
%
%
%
\section{Quantum speed limit time for arbitrary initial state}\label{Sec3}
Let's assume that the quantum system initially is in a single-qubit state  $\rho_{0} = \frac{1}{2} (\mathbf{I} + \sum_{i=1}^{3} r_{i} \sigma_{i})$,  where  $\mathbf{I}$ is the identity operator , $\sigma_{i}$'s ($i=1,2,3$) are the Pauli operators, and $r_{i}$'s are the components of Bloch vector. At time $t$ the evaluated state of the open quantum system is represented by $\rho_{t}$. The dynamics of such a quantum system can be described by the time-dependent master equations of the form $\dot{\rho_{t}}=\mathcal{L}_{t}(\rho_{t})$, where $\mathcal{L}_{t}$ is the generator of the evolution \cite{Breuer}. Now we want to calculate the minimum time  it takes for a system to evolve from state $\rho_{\tau}$ to state $\rho_{\tau+\tau_{D}}$, where $\tau_{D}$ is the driving time. This minimum time is called QSL time $\tau_{QSL}$. One should use a suitable distance measure to characterize QSL time . In Ref. \cite{Zhang} Zhang et al. have used relative purity as the distance measure to quantify QSL time $\tau_{QSL}$. The important point about their QSL time is that, it can be used for both mixed and pure initial states.  Relative purity $\mathcal{R}(\tau)$ between initial state $\rho_{\tau}$ and evolved state $\rho_{\tau + \tau_D}$ can be written as $\mathcal{R}(\tau + \tau_D)=tr(\rho_{\tau}\rho_{\tau + \tau_D})/tr(\rho_{\tau}^{2})$. Following the methodology presented in Ref. \cite{Zhang}, one can obtain the ML bound of QSL time for dynamics of open quantum system as 

\begin{equation}\label{ML}
\tau \geq \frac{\vert \mathcal{R}( \tau + \tau_D ) -1 \vert tr (\rho_{\tau}^{2})}{\overline{ \sum_{i=1}^{n} \sigma_{i} \rho_{i}}},
\end{equation}

where $\sigma_{i}$ and $\rho_{i}$ are the singular values of $\dot{\rho_{t}}$ and $\rho_{\tau}$, respectively, $\overline{\mathcal{B}}=\frac{1}{\tau_{D}} \int_{\tau}^{\tau + \tau_{D}} \mathcal{B} dt$. Doing an analogous procedure,  MT bound of QSL time for dynamics of open quantum system can be derive as
\begin{equation}\label{MT}
\tau \geq \frac{\vert \mathcal{R}( \tau + \tau_D ) -1 \vert tr (\rho_{\tau}^{2})}{\overline{ \sqrt{\sum_{i=1}^{n} \sigma_{i}^{2}}}}.
\end{equation}
Considering these two bound, one can define the unified bound as
\begin{equation}\label{QSLT}
\tau_{QSL}=\max \lbrace \frac{1}{\overline{ \sum_{i=1}^{n} \sigma_{i} \rho_{i}}}, \frac{1}{\overline{\sqrt{ \sum_{i=1}^{n} \sigma_{i}^{2}}}} \rbrace \times \vert \mathcal{R}( \tau + \tau_D ) -1 \vert tr (\rho_{\tau}^{2}).
\end{equation}
For dephasing coupling, QSL time for dynamics of quantum system $\mathcal{S}$ can be written as \cite{Zhang}
\begin{equation}
\tau_{QSL}=\frac{C_{l_1}(\rho_{\mathcal{S}}(0))\vert f(\tau)f(\tau+\tau_D)-f^{2}(\tau) \vert}{\frac{1}{\tau_{D}}\int_{\tau}^{\tau+\tau_D} \vert \dot{f}(t)\vert dt},
\end{equation}
where $C_{l_1}(\rho_{\mathcal{S}}(0))=\sqrt{r_1^{2}+r_2^2}$ is the coherence of initial state. Hereafter we consider the initial maximal coherent state with Bloch vector $(r_1=r_2=1/\sqrt{2},r_3=0)$. In Fig. \ref{Fig4}(a), QSL time is plotted as a function of initial time parameter $\tau$ for dephasing model with sub-Ohmic environment $s=0.5$ and driving time $\tau_{D}=1$. As can be seen in the  presence of disturbance effects  QSL time is shorter than  QSL time when there exist no disturbance effects. Actually, It is due to the fact that the existence of disturbance leads to non-Markovian quantum evolution. So the quantum evolution is faster than the case in which the environment is in  equilibrium.
\begin{figure}[!] 
\centerline{\includegraphics[scale=0.7]{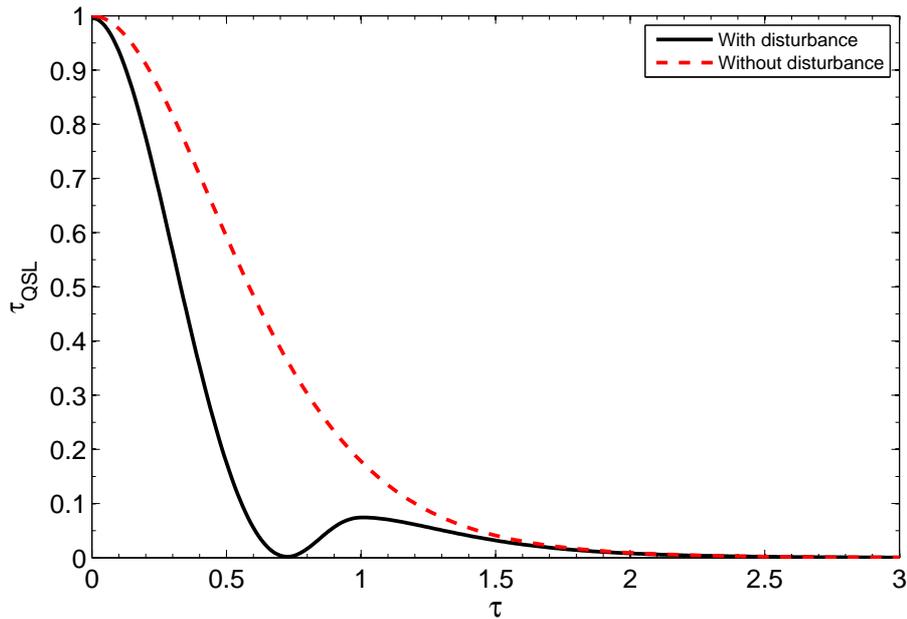}}
\vspace*{8pt}
\caption{(color online) QSL time as a function of initial time parameter $\tau$ for dephasing model with Lorentzian spectral density with $\gamma=10$ and $\Delta/\lambda=1$ with driving time $\tau_{D}=1$. Black line shows the QSL time in the presence of disturbance $\langle \sigma_{z}^{a} \rangle=0.05$ and Red dashed line shows the QSL time without disturbance.}\label{Fig5}
\end{figure}
In Fig. \ref{Fig4}(b), QSL time is plotted as a function of initial time parameter $\tau$ for dephasing model with Ohmic environment $s=1$ and driving time $\tau_{D}=1$. As can be seen in the  presence of disturbance effects  QSL time is shorter than  QSL time when there exist no disturbance effects. Actually, It is due to the fact that existence of disturbance leads to non-Markovian quantum evolution, hence the quantum evolution is faster than the case in which the environment be in  equilibrium.
In Fig. \ref{Fig4}(c), QSL time is plotted as a function of initial time parameter $\tau$ for dephasing model with super-Ohmic environment $s=2$ and driving time $\tau_{D}=1$. As can be seen in the  presence of disturbance effects  QSL time is shorter than  QSL time when there exist no disturbance effects. Actually, It is due to the fact that existence of disturbance leads to non-Markovian quantum evolution, hence the quantum evolution is faster than the case in which the environment be in  equilibrium.
In Fig. \ref{Fig5}, QSL time is plotted as a function of initial time parameter $\tau$ for dephasing model with Lorentzian spectral density. In order to show the effects of disturbance purely we choose $\gamma=10$ and $\Delta/\lambda=1$. As can be seen in the  presence of disturbance effects  QSL time is shorter than  QSL time when there exist no disturbance effects. Actually, It is due to the fact that existence of disturbance leads to non-Markovian quantum evolution, hence the quantum evolution is faster than the case in which the environment be in  equilibrium.
\section{Summary and conclusion}\label{Sec4}
In this work we considered the dephasing model in which two quantum systems $\mathcal{A}$ and $\mathcal{S}$ interacts with environment sequentially. The environment is disturbed by the interaction of first quantum system $\mathcal{A}$ with environment. Then the quantum system $\mathcal{S}$ interacts with environment which has been disturbed. Note that, If the time interval between the beginning of the interaction of quantum system $\mathcal{S}$ and the end of the interaction of quantum system $\mathcal{A}$ with environment is greater than correlation time of the environment then environment returns to the  equilibrium before interacting with the quantum system $\mathcal{S}$. According to Eq. \ref{coherence}, one can concluded that parameter $\langle \sigma_{z}^{a} \rangle$ defines the amount of disturbance in environment. In general, the coherence of the firs quantum system $\mathcal{A}$ quantify the power of disturbance of the environment. In the sense  that if initial state of the quantum system $\mathcal{A}$ is maximally coherent then disturbance has its maximum value and  and vice versa \cite{Masashi1}.
 We studied the effects of disturbance on the QSL time. Here,  two types of environment  were considered with Ohmic and  Lorentzian spectral density. In this work, We have shown that the disturbance of the environment leads to the non-Markovianity of the quantum evolution  of  quantum system $\mathcal{S}$. In Ref. \cite{Deffner1}, authors have shown that non-Markovian effects reduce the QSL time. In confirmation of their result, we showed that in the presence of disturbance effects the QSL time is shorter than the case in which there is no disturbance.

\end{document}